\documentclass[10pt]{iopart} %for two columns
\usepackage{graphicx}% Include figure files
\usepackage{dcolumn}% Align table columns on decimal point
\usepackage{bm}% bold math
\usepackage{color}

\newcommand{\Tczero}{$T_{\mathrm{c}}^{\mathrm{zero}}$}
\newcommand{\Tconset}{$T_{\mathrm{c}}^{\mathrm{onset}}$ }
\newcommand{\RH}{$R_{\mathrm{H}}$ }

\newcommand{\Tc}{$T_{\mathrm{c}}$}

\newcommand{\1}{sample \#1 (4 nm)}
\newcommand{\2}{sample \#2 (19 nm)}
\newcommand{\B}{$B$} 
\usepackage{ulem}
\newcommand{\add}[1]{\textcolor{black}{#1}}	%←追加
\newcommand{\erase}[1]{\if0{#1}\fi}	%←追加
\begin{document}

\paper{Anisotropy of upper critical fields and interface superconductivity in FeSe/SrTiO$_3$ grown by PLD}

\author{Tomoki Kobayashi, Hiroki Nakagawa, Hiroki Ogawa, Fuyuki Nabeshima, and Atsutaka Maeda}

\address{Dept. of Basic Science, University of Tokyo, Meguro, 153-8902, Tokyo, Japan}
\ead{kobayashi-tomoki375@g.ecc.u-tokyo.ac.jp}

\begin{abstract}
In this study, we grow FeSe/SrTiO$_{3}$ with thicknesses of 4--19 nm using pulsed laser deposition and investigate their magneto-transport properties.
The thinnest film (4 nm) exhibit negative Hall effect, indicating electron transfer into FeSe from the SrTiO$_{3}$ substrate.
This is in agreement with reports on ultrathin FeSe/SrTiO$_{3}$ grown by molecular beam epitaxy.
The upper critical field is found to exhibit large anisotropy ($\gamma >$ 11.9), estimated from the data near the transition temperature (\Tc).
In particular, the estimated coherence lengths in the perpendicular direction are 0.15--0.27 nm, which are smaller than the c-axis length of FeSe, and are found to be almost independent of the total thicknesses of the films.
These results indicate that superconductivity is confined at the interface of FeSe/SrTiO$_{3}$.

\end{abstract}

\vspace{2pc}
%\noindent{\it Keywords\/}: iron-based superconductor, PLD grown films, anisotropy of upper critical fields, interface superconductivity 
%\keywords{}

%\submitto{\JPCM}
\maketitle
\ioptwocol

\section{Introduction}

Monolayer FeSe on SrTiO$_{3}$ (STO) shows a significantly enhanced superconducting transition temperature (\Tc) \cite{Wang_2012,He_2013} from that of 9 K in the bulk form \cite{Hsu_2008}, which was claimed by the 
opening of the superconducting gap-like structure at a temperature as high as 65 K observed by the angle-resolved photoemission spectroscopy (ARPES) measurements \cite{Wang_2012,He_2013}.
The \Tc \  enhancement is mainly attributed to the interfacial electron-phonon coupling with optical phonons of STO \cite{Liu_2012} and/or electron transfer from the STO substrate \cite{Lee_2014}.
However, zero-resistivity has been reported only below 30 K by transport measurements \cite{Wang_2012, Pedersen_2020, Faeth_2021}, except for the singular report of \Tc\ $>$ 100 K \cite{Ge_2015}.
Although the large difference between the spectroscopic \Tc\  and the transport \Tc \ has been discussed in terms of the superconducting fluctuation in two-dimensional systems\cite{Faeth_2021}, zero resistivity below 46 K is realized even in surface-electron-doped FeSe films by the electric-double-layer transistor (EDLT) technique \cite{Kouno_2018,Shikama_2020,Shikama_2021}, suggesting the presence of further space for the improvement of transport \Tc\ of the interface superconductivity in FeSe/STO.
For that purpose, systematic investigation of the interface structure for a wide range of combinations with oxide materials must be necessary, and we believe that pulsed laser deposition (PLD) is a suitable growth technique.
Indeed, we fabricated FeSe films on the STO substrate whose surface was atomically flat and found that the interface superconductivity was realized even in the PLD-grown films, based on the strain vs. \Tc \  relation and the film thickness vs. \Tc \ relation \cite{Kobayashi_2022}.
However, stronger evidence of the interface superconductivity in the PLD-grown films is desired.

In this study, we grow FeSe/STO with thicknesses of 4--19 nm using PLD and investigated their magneto-transport properties.
The thinnest film (4 nm) exhibits negative Hall effect, indicating electron transfer into FeSe from the STO substrate.
This is in agreement with reports on ultrathin FeSe/STO grown by molecular beam epitaxy (MBE) \cite{Wang_2015,Zhao_2018}.
The upper critical field estimated from the data near \Tc\  was found to exhibit large anisotropy ($\gamma\sim$ 11 -- 25).
In particular, the estimated coherence lengths in the perpendicular direction are 0.15--0.27 nm, which is smaller than the c-axis length of FeSe, and are almost independent of the total thickness of the film.
These results indicate that superconductivity is confined to the interface of FeSe/STO.

\section{Experimental methods}
FeSe films were grown on atomically flat STO (100) substrates using the PLD technique\cite{Kobayashi_2022}.
The preparation of an atomically flat surface of the STO substrate is essential for realizing the interface superconductivity in this study.
Ultrathin films of FeSe deteriorate easily in air, leading to the disappearance of superconductivity.
In order to protect the films from degradation caused by the air-exposure, we deposited amorphous Si on FeSe/STO at room temperature also by PLD.
In order to obtain electrical contact with the Si-deposited FeSe/STO, FeSe was deposited as electrodes at room temperature prior to the growth of the crystalline FeSe film.
The orientation and the crystal structure of the grown films were characterized using
X-ray diffraction (XRD) measurements with Cu K$\alpha$ radiation at room temperature.
The thicknesses of thick films were estimated by the Laue fringes of the (001) reflection of FeSe.
As for the ultrathin films, where FeSe reflections were not observed due to weak intensities, the film thicknesses were estimated by the growth rate, which is always monitored for all films including thicker ones.
Resistivity and Hall measurements were conducted using a physical property measurement system (PPMS) from 2 to 300 K.

\section{Results and discussion}
\begin{figure}[htbp]
%\centering 

\includegraphics[width=\linewidth]{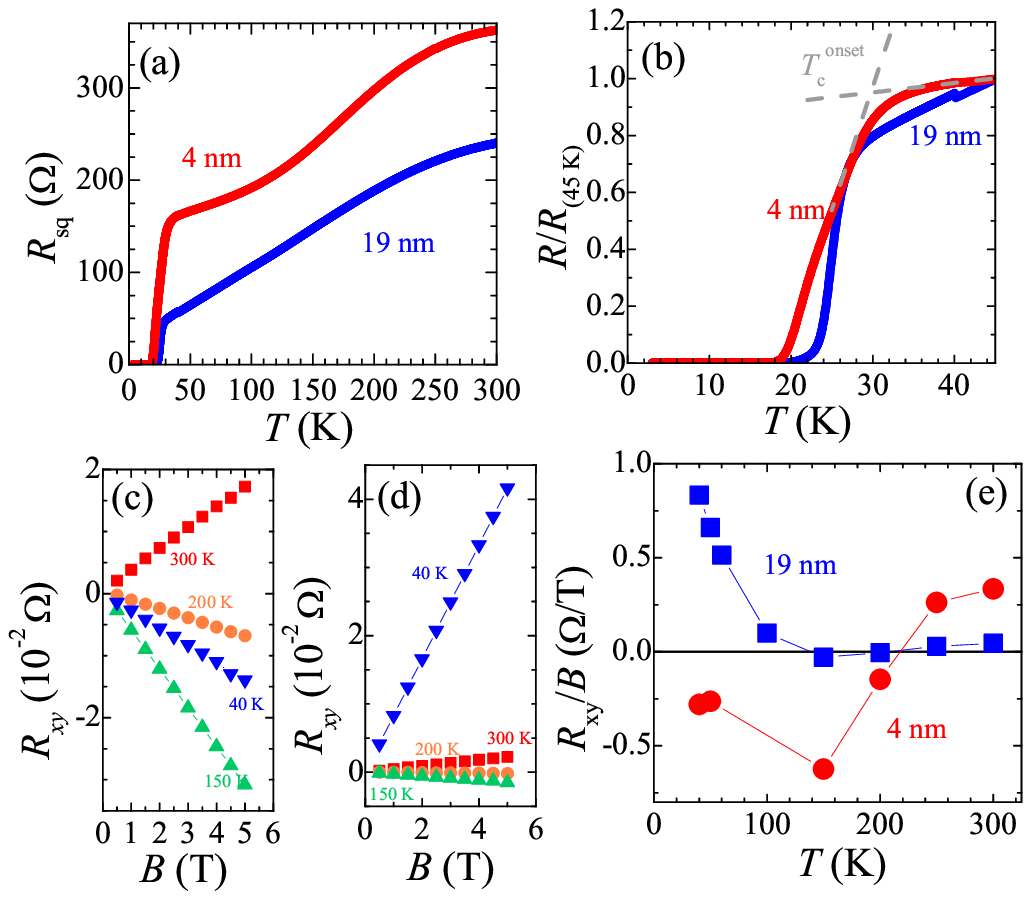}
\caption{(a) Temperature dependence of the sheet resistance of \1 and \2, and (b) the expanded plot in the vicinity of $T_c$.  The sheet resistance is normalized by that at 45 K. (c) Magnetic-field dependence of Hall resistance $R_{xy}$ of \1 at different temperatures and (d) those for \2. (e) temperature dependence of $R_{xy}/B$ of \1 and \2.  $R_{xy}/B$ is equal to $R_H/d$, where $R_H$ and $d$ are the Hall coefficient and the film thickness, respectively.}
\label{fig:RT} 
\end{figure}

Figures \ref{fig:RT}(a) and \ref{fig:RT}(b) show the temperature dependence of the sheet resistance of samples \#1 and \#2 with thicknesses of 4 and 19 nm, respectively.
Both samples showed superconductivity at low temperatures.
In the previous study using PLD \cite{Kobayashi_2022}, even a 5-nm-thick film did not show superconductivity possibly because of the degradation by the air exposure. 
The appearance of superconductivity in \1 indicates that the capping layer of Si successfully protects the sample from the air exposure.
The onset \Tc\  (\Tconset) of the thinner \1 was 30 K and higher than that of \2 (\Tconset = 26 K).
This is in agreement with our previous results\cite{Kobayashi_2022} and is a common trend among ultrathin FeSe/STO including those grown by MBE\cite{Wang_2015,Zhao_2018}, which is considered to be one of the hallmarks of the interface superconductivity.
\add{
One might think that superconductivity properties are the same between the thinner film and the thicker film if superconductivity takes place only at the interface, which is different from our observation.
We consider the origin of the difference to be due to the difference in the carrier density near the interface.
We will discuss this issue later again.}
\erase{However, the}
The transition of \1 was broader than that of \2, \erase{indicating} which indicates more inhomogeneity in the thinner film.
\add{We consider the reason for the observed more inhomogeneous behavior in the thinner film as follows. 
We deposited amorphous Si on the FeSe surface for the protection of FeSe by the PLD technique.  Since amorphous Si deposited by the PLD has high kinetic energy\cite{Willmot_2000}, this can damage an area down to a few nm from the FeSe surface.  Thus thinner films can be more strongly damaged, leading to the observed behavior.}
%Figure-----------
%------------

Figures \ref{fig:RT}(c)--(e) show the temperature dependence of the Hall resistance $R_{xy}$ as a function of magnetic field $B$.
As shown in Fig. \ref{fig:RT}(c) and \ref{fig:RT}(d), both films showed linear dependence of Hall resistance $R_{\mathrm{xy}}$ on magnetic fields from 300 K to 40 K.
In particular, in \1, $R_{\mathrm{xy}}$ showed a negative gradient with $B$ below 200 K, while it became positive at room temperature (Fig. \ref{fig:RT}(c)), which can be recognized more easily in Fig.\ref{fig:RT}(e), where $R_{\mathrm{xy}}/B$ (which is equal to $R_{\mathrm{H}}/d$, where \RH and $d$ is the Hall coefficient and the film thickness, respectively) are plotted as a function of temperature.
This is in contrast to the Hall data in FeSe flakes\cite{Farrar_2020} and thin films grown on substrates without any special treatment\cite {Nabeshima_2018}, where \RH shows a positive sign even below 150 K.
Our result for \1 is rather similar to the MBE-grown FeSe/STO where \RH exhibits negative signs at low temperatures, exhibiting the electron transfer into FeSe from the STO substrate\cite{Wang_2015,Zhao_2018}.
Thus, the above results showed that the carrier transfer from the STO substrate takes place even in the PLD-grown FeSe/STO films as well as in the MBE-grown films.
In contrast, \2 shows the positive $R_{\mathrm{xy}}/B$ even at low temperatures, and the temperature dependence of $R_{\mathrm{xy}}/B$ is similar to those of FeSe films without the substrate treatment\cite{Nabeshima_2018}, although it shows "high" \Tc \ which is characteristic of the interface superconductivity.
The difference in $R_{\mathrm{xy}}/B$ between \1 and \2 can be understood as follows.
\erase{Electron transfer is limited to a few layers near the FeSe/STO interface so that the }
\\
\erase{carriers in the upper layers provide positive numbers for the Hall effect.}\\
\add{
Electron transfer is limited to a few layers near the FeSe/STO interface in \1, whereas in the thicker sample \2 transferred carrier is diluted, leading to the positive $R_{\mathrm{xy}}/B$ as a whole.
Thus, thicker films showed positive $R_{\mathrm{xy}}/B$ even at low temperatures.
This also explains why superconductivity properties are different between the two samples even when superconductivity takes place at the interface.
The difference in the carrier density can cause the difference in superconductivity properties.
Indeed, a similar discussion is found in a paper\cite {Zhao_2018}, where they investigated a series of FeSe/STO films with different thicknesses.
In addition to the transport data such as resistivity and Hall effects, the authors used atomically resolved electron energy-loss spectroscopy (EELS) and concluded that the superconductivity is confined at the interface regardless of FeSe thickness, although \Tc\ differs due to the difference in the number of electrons at the interface. 
}
%Figure-----------
\begin{figure}[htbp]
\centering 
\includegraphics[width=\linewidth]{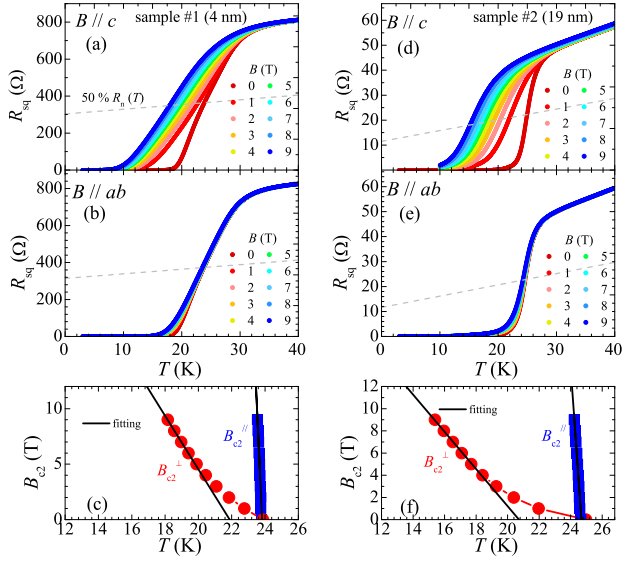}

\caption{(a),(b) Temperature dependence of sheet resistance of \1 for \B$/\!/\textit{c}$ and \B$/\!/\textit{ab}$, respectively. (c) Temperature dependence of the upper critical fields of \1 for \B$/\!/\textit{c}$, $B_{c2}^{\perp}$ (red) and \B$/\!/\textit{ab}$, $B_{c2}^{/\!/}$ (blue).
(d),(e) Temperature dependence of sheet resistance of \2 for \B$/\!/\textit{c}$ and \B$/\!/\textit{ab}$, respectively. (f) Temperature dependence of $B_{c2}^{\perp}$ (red) and $B_{c2}^{/\!/}$ (blue) of \2.
}
\label{fig:fig2} 
\end{figure}
%---------------

To investigate the anisotropy of superconductivity in the grown films, we conducted magneto-transport measurements on the grown films across \Tc.
Figure \ref{fig:fig2} shows the temperature dependence of the sheet resistance of the films under magnetic fields up to 9 T for \B$/\!/\textit{c}$ and \B$/\!/\textit{ab}$.
For \B$/\!/\textit{c}$, \Tconset of \1 decreases slightly from 26 K to 20 K, and the superconducting transition became broader with increasing field (Fig. \ref{fig:fig2}(a)).
One the other hand, for \B$/\!/\textit{ab}$ (Fig. \ref{fig:fig2}(b)), \Tconset was not changed almost at all, and a little broadening was observed near \Tczero, suggesting a very high upper critical field $B_{\mathrm{c2}}$.
%These changes of superconducting transition under magnetic fields.
%\cite{Sun_2014,Wang_2015,Liu_2020}.
$B_{\mathrm{c2}}(T)$ for \B$/\!/\textit{c}$ ($B_{c2}^{\perp}(T)$) and \B$/\!/\textit{ab}$ ($B_{c2}^{/\!/}(T)$)  was determined by taking a criterion of the field at 50\% of the normal-state resistance $R_{\mathrm{n}}(T)$ (Fig.\ref{fig:fig2}(c)).
With the aid of the Werthamer–Helfand–Hohenberg (WHH) formula \textit{i.e.} $B_{c2}(0)=-0.69 T_{\mathrm{c}}\ |dB_{\mathrm{c2}}/dT_{\mathrm{c}}|$, where the only orbital pair breaking effect is considered, and setting \Tc\ = \Tconset, we obtained $B_{c2}(0)$.
Note that the paramagnetic effect and the spin-orbital coupling should be considered to estimate actual $B_{c2}(0)$.
Indeed, even in an early publication on bulk single crystals of FeSe, the upper critical field becomes almost isotropic at low temperatures, which is considered to be due to the paramagnetic pair braking\cite{Vedeneev_2013}.
However, the above analysis is more suitable to discuss anisotropiy of the coherence length of superconductivity origin.
We obtained $B_{c2}^{\perp}(0)=45.8$ T and $B_{c2}^{/\!/}(0)$=675 T for \1, providing a very large anisotropy parameter $\gamma=B_{c2}^{/\!/}(0)/B_{c2}^{\perp}(0)=14.8$.
Similar data for \2 shown in Fig. \ref{fig:fig2}(d--f) and the same analysis also exhibited a large anisotropy with $\gamma=11.9$.
These values of $\gamma$ are larger than those of much thicker FeSe films grown on the substrate without any special treatment ($\gamma \sim 2.5)$)\cite{Sawada_2016} and those of intercalated FeSe systems with an expanded distance between superconducting FeSe layers ($\gamma \le 11)$)\cite{Wang_2016,Hanisch_2020}.
The GL coherence lengths of \1 are $\xi_{ab} (0)=\sqrt{\phi_{0}/2\pi B_{c2}^{\perp}}$ = 2.68 \AA\ and $\xi_{c}(0)=\sqrt{\phi_{0}/2\pi B_{c2}^{/\!/}\xi_{ab}}$ = 0.18 \AA.
The obtained $\xi_{ab} (0)$ is similar to that of monolayer FeSe/STO grown by MBE\cite{Fan_2015}. 
Interestingly, the estimated $\xi_{c}(0)$ is shorter than the distance between FeSe layers, suggesting superconductivity is confined in very thin layers.

\begin{figure}[htbp]
\begin{center}
\includegraphics[width=\linewidth]{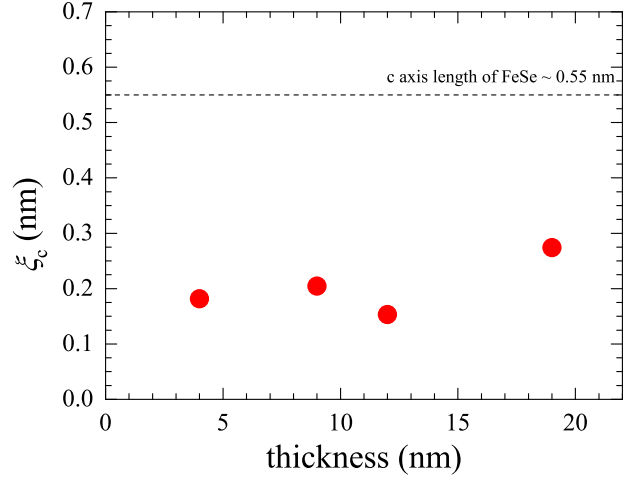}
\caption{The GL coherence lengths $\xi_\mathrm{c}(0)$ as a function of the film thckness.}
\end{center}
\label{fig:fig3} 
\end{figure}

To look for the origin of the extremely confined superconductivity, we investigated $\xi_{c} (0)$ for various films with different thicknesses (Fig. \ref{fig:fig3}).
$\xi_c(0)$ were 0.15--0.27 nm, which are also shorter than the distance between FeSe layers, and were found to show a very weak dependence on the film thickness. This result suggests that the presence of the FeSe/STO interface is essential and that the superconductivity is confined at the FeSe/STO interface.
%Figure-----------
%---------------
It is suggestive to compare our result with those of thin flakes of FeSe\cite{Farrar_2020}, where $\xi_c$ was investigated for the thickness in the range 9--100 nm.
For thick flakes, $xi_c$ was $\sim$ 1--2 nm, whereas it decreases down to $\sim$0.6 nm with decreasing the thickness of the flake, which the authors of ref.\cite{Farrar_2020} claimed was evidence that superconductivity is confined to one FeSe layer in the thin-flake limit.
On the other hand, in our thin films, $\xi_c\sim$0.15--0.27 nm, which is smaller than FeSe-FeSe distance and is nearly independent of the total film thickness.  This suggests that the superconductivity observed in our ultrathin films is qualitatively different from those observed in the thin flakes, and again suggests that it is confined at the interface between FeSe and the STO substrate.

Therefore, together with a negative Hall effect in thinner films, the above-mentioned very short $\xi_\mathrm{c}(0)$ which was almost independent of the total film thickness is additional evidence that the interface superconductivity is realized in our PLD-grown films.

\section{Conclusion}
In conclusion, we grew FeSe/STO with different thicknesses of 4--19 nm using PLD and investigated the magneto-transport effect, namely the Hall effect and the the upper critical fields for \B$/\!/\textit{c}$ and \B/$\!$/\textit{ab}.
The thinnest film (4 nm thickness) showed negative $R_{xy}/d$ even down to low temperatures, indicating the electron transfer from the STO substrate.
The anisotropy of the upper critical fields estimated from the data near \Tc\ was very large ($\gamma \ge$ 11.9), and the estimated c-axis coherence length, $\xi_{c}(0)$ is very short ($\le$ 0.27 \AA) and almost independent of the total film thickness.
All of these results indicate that superconductivity is confined at the interface of FeSe/STO in our PLD-grown films.
Thus, the next step is to realize higher zero-resistance \Tc\  of this interface superconductivity, which is under investigation.

\begin{ack}
We would like to acknowledge K. Ueno at the University of Tokyo for their technical support of the XRD measurements.
\end{ack}

%\section*{Data Availability Statement}
%The data that support the findings of this study are available from the corresponding author upon reasonable request.
\section*{References}
\bibliography{manuscript_JPCM_revised}% Produces the
\bibliographystyle{iopart-num}
\end{document}